# Liability-side Pricing of Swaps and Coherent CVA and FVA by Regression/Simulation

Lou Wujiang[1]
1st Draft, August 8, 2014. Updated Dec 20, 2015.

**Abstract**
An uncollateralized swap hedged back-to-back by a CCP swap is used to introduce FVA. The open IR01 of FVA, however, is a sure sign of risk not being fully hedged, a theoretical no-arbitrage pricing concern, and a bait to lure market risk capital, a practical business concern. By dynamically trading the CCP swap, with the liability-side counterparty provides counterparty exposure hedge and swap funding, we find that the uncollateralized swap can be fully replicated, leaving out no IR01 leakage. The fair value of the swap is obtained by applying to swap's net cash flows a discount rate switching to counterparty's bond curve if the swap is a local asset or one's own curve if a liability, and the total valuation adjustment is the present value of cost of funding the risk-free price discounted at the same switching rate. FVA is redefined as a liquidity or funding basis component of total valuation adjustment, coherent with CVA, the default risk component. A Longstaff-Schwartz style least-square regression and simulation is introduced to compute the recursive fair value and adjustments. A separately developed finite difference scheme is used to test and find regression necessary to decouple the discount rate switch. Preliminary results show the impact of counterparty risk to swap hedge ratios, swap bid/ask spreads, and valuation adjustments, and considerable errors of calculating CVA by discounting cash flow or potential future exposure.

## 1. Introduction

Counterparty credit risk of swaps is well studied in both pricing (Duffie and Huang 1996) and econometrics (Duffie and Singleton 1997). Pre-crisis studies, however, focus on the default risk of the swap counterparties (Collin-Dufresne and Solnik 2001, and Liu et al 2006), as derivatives financing cost only manifests its impact during the 2007-2009 global financial crisis. FVA (funding valuation adjustment) has since become the main thread of a broader subject of funding cost's impact on derivatives pricing and valuation. In particular, following the industrywide adoption of OIS discounting for fully collateralized derivatives, uncollateralized trades pricing and valuation adjustments prove to be a much controversial and involving problem.

A common theme when introducing FVA of an uncollateralized customer swap is to consider a back-to-back swap with a fully collateralized swap clearing facility (CCP, Central Counterparty) member firm. The CCP swap is of the exact terms and notional as the customer swap and is assumed to have hedged the former so that the net economic effect would be that of making a loan with the customer swap as collateral. A bank, for instance, has to post cash-like collateral to the CCP when the customer swap is in-the-money (ITM), and the financing cost of the collateral is defined as FVA, here a cost thus FCA (funding cost adjustment). An FVA benefit occurs when the swap is out-of-the-money (OTM) as the bank gets cash from the CCP.

---

[1] The views and opinions expressed herein are the views and opinions of the author, and do not reflect those of his employer and any of its affiliates. The author wishes to thank Fabio Mercurio.

While the back-to-back CCP swap serves as a perfect cash flow hedge, it is nevertheless imperfect from the risk point of view, as FVA itself bears unhedged interest rate sensitivity. Green and Kenyon (2014), for example, calculates the open IR01, which would attract market risk capital[2]. The uncollateralized swap is not fully replicated by the CCP swap and the FVA-adjusted fair value cannot be claimed arbitrage-free. In practice, counterparty default aware payoffs are specified in an ad-hoc manner and discounted at the risk free rate to arrive at various valuation adjustments. There is however no in-depth discussion on whether the risk-free price when added with all valuation adjustments is indeed arbitrage-free (Bielecki and Rutkowski 2013).

Burgard and Kjaer (2011, 2013) present a rigorous micro-economic analysis of an uncollateralized option dynamically hedged with underlying stocks and both counterparties' debt instruments. Because the funding arrangement and own default risk hedge are made from the dealer's own perspective, the resulted fair value violates the law of one price (LOP), a criteria of no-arbitrage pricing, and thus can only be viewed as private value (Hull and White 2012). Green and Kenyon (2014) take the private valuation approach a step further by incorporating a regulatory capital charge into the fair value, and later argue that the risk neutral pricing principle no longer applies (Kenyon and Green 2014).

In a departure from this stream of literature, Lou (2015a) allows the party on the liability side to hedge its counterparty's exposure by depositing cash outside of the derivatives netting set with the credit risk mitigation effect realized by the Set-Off provision of the ISDA Master Agreement or by common laws. The deposit also becomes a source of funding for the derivative. This arrangement is economically neutral to both parties if the cash deposit earns the same interest rate as the market rate of the depositor's debt, and it avoids the often controversial use of selling protection on self or dealing in own debts for the purposes of hedging one's own default and funding risk. An uncollateralized option is attainable, the option fair value V is arbitrage free, and risk neutral pricing still applies, so long as the risk neutral world covers the corporate bond markets.

The liability-side pricing (LSP) theory developed therein represents the first no-arbitrage pricing model that neatly integrates pricing of market risk, counterparty credit risk, and funding risk. It eliminates funding arbitrage opportunity, e.g., signing a full CSA or no CSA. Valuation adjustments can now be defined rigorously and consistently as price differences of the same model when input funding and credit curves are shifted. The so called coherent CVA and FVA add up to the precisely defined total counterparty risk adjustment (CRA), each corresponding to default risk and funding (or liquidity) risk. In the context of the un-concluding FVA debate, the coherent CVA and FVA definition offers advantages such as IFRS 13 compliance, observation of the law of one price, and no double counting of DVA and FVA.

---

[2] Residue IR01 normally get an add-on market risk capital requirement in proportion to IR01.

The theory's application to swaps needs justification, as swaps don't permit the kind of dynamic trading in underlying as stock options. This paper's first contribution is to extend the liability-side pricing theory to uncollateralized swaps by setting up a dynamic hedge in CCP swaps where the notional of the swap hedge changes so that the uncollateralized swap's interest rate sensitivity is fully hedged. This results in a partial differential equation (PDE) similar to the extended Black-Scholes-Merton PDE for options and proves that an uncollateralized swap can be dynamically replicated by CCP swaps. The open IR01 is plugged in and the same coherent CVA and FVA then applies to swaps.

As is with most FVA and XVA formulations, coherent CVA and FVA is recursive in that the effective discount rate switches on the fair value itself. Finite difference (FD) schemes (Lou 2015b) are subject to the curse of dimensionality and ineffective at netting set levels. This paper's second contribution is to devise a least-square regression, Monte Carlo simulation scheme for the liability-side pricing model to decouple the recursion and to efficiently compute fair value, CVA and FVA, with numerical results verified by a FD solver.

This paper proceeds with Section 2 briefly reviewing the main results of the liability-side pricing and conducting dynamic replication of Non-CCP swaps with CCP swaps. Section 3 then gives the key result of swap valuation in Proposition 1 with a proof in the Appendix and valuation adjustments. In section 4 we cover numerical techniques and build a mixed-normal-lognormal short rate model and the Black-Karasinski short rate model for the LIBOR rate to evaluate the impact of counterparty default risk and funding basis in the current low rate environment on swap pricing and valuation adjustments. Numerical results are presented in Section 5 and Section 6 concludes.

## 2. Replicating an Uncollateralized Swaplet with a CCP Swaplet

To begin, let's review the main results of the liability-side pricing of uncollateralized stock option products. The option fair value V is governed by a modified Black-Scholes-Merton PDE with the option's instantaneous rate of return at its liability-side's rate,

$$\frac{\partial V}{\partial t} + (r_s - q)S\frac{\partial V}{\partial S} + \tfrac{1}{2}\sigma^2 S^2 \frac{\partial^2 V}{\partial S^2} - r_e V = 0$$

where $r_e$ is the effective discount rate switching between two counterparties' rates $r_b$ and $r_c$, $r_e = r_b I(V(t) < 0) + r_c I(V(t) \geq 0)$, $r_s$ the stock repo rate, $q$ dividend yield, $\sigma$ volatility, and $S$ stock price. Equivalently, for an option expiring at $T$ with payoff function $H(T)$, the risk neutral pricing principle or formula is extended as follows,

$$V(t) = E_t^Q [e^{-\int_t^T r_e du} H(T)],$$

Q is an equivalent martingle measure in a risk-neutral world that consists of the stock, the risk-free bank account, and the corporate bonds of the parties.

The dynamic hedging exercise employed to arrive at the above, however, works only for derivatives of tradeable underlying instruments. For interest rate swaps and derivatives, the underlying rate is not a tradeable instrument. Thanks to post-crisis market development, we could use CCP swaps as the hedging instruments for uncollateralized swaps. Indeed, FVA has been commonly introduced on a back-to-back hedge of an uncollateralized swap with a CCP swap, i.e., a static hedge. Here what we do differently is to have a dynamic hedging strategy to replicate an uncollateralized swap with a CCP swap and to extend the liability-side pricing principle to swaps.

Consider a hypothetical dealer or bank (party B) and an uncollateralized customer (party C) enter into an interest rate swap (swap #1). The bank hedges the interest rate risk by continuously trading in an identical swap (swap #2) with a CCP member firm. Swap #1 has a unit notional and swap #2 a dynamic notional of $\Delta_t$. $\Delta_t$ becomes the swap hedge ratio. Back-to-back hedge is a simple case with $\Delta_t =1$ for all $t<=T$.

Swap #2 is fully collateralized in cash and is priced at the risk-free discount curve. Let $V^*$ denote the fair value of swap #2 per unit notional, $L_s$ the collateral account balance under the CCP swap clearing agreement[3], $r_L$ is the interest rate paid on $L_s$, then $L_s=\Delta_t V^*_t$, assuming that collaterals are maintained perfectly and continuously.

For the time being, suppose swap #1 is a single reset (at $T-\Delta T$), single payment (at $T$) swap or swaplet. Let $L_t$ denote the mutually funded cash deposit balance. Write $L_t = L_t^+ - L_t^-$, $L_t^+$ the cash amount posted (deposited) by party C to B that pays C's cash debt interest rate $r_c(t)$, and $L_t^-$ the cash collateral posted by B to C earning B's interest rate $r_b(t)$. Furthermore, the fair value of swap #1 is fully covered by the deposit[4], i.e., $L=V$.

$M_t >=0$ is the bank account balance that earns the cash deposit rate $r$. $N_t>=0$ is B's short term debt (borrowing) account balance that pays par rate $r_N(t)$, $r_N(t)>=r(t)$. The wealth equation of the hedged swap economy reduces to the balance of the bank account and the debt account,

$$\pi_t = M_t + (1-\Gamma_t)(V_t - L_t - N_t - \Delta_t V^*_t + L^s) = M_t - (1-\Gamma_t)N_t,$$

where $1-\Gamma$ is party B and C's joint survival indicator. To fully replicate the pre-default portfolio, we set $\pi=0$. Consequently both $M_t$ and $N_t$ are set to zero, trivially.

If the customer swaplet is a receivable to the bank, i.e., $V_t>=0$, the customer is indifferent to making a cash deposit to the bank in the same amount so long as the deposit earns its current market debt rate. Assuming a functioning debt market, the customer could raise that amount of the cash. Now if the bank defaults, the cash deposit as a receivable to C and the derivative as a payable to C set off and there is no default settlement. The same is true if C is default. From B's perspective, the customer swaplet is financed by C's deposit and its exposure to C is also hedged, see detailed arguments in Lou 2015a.

---

[3] CCP default risk is secondary to no-CCP entities so we assume CCP non-defaultable in this exercise. CCP's initial margin requirement related funding cost is considered separately.

[4] Any upfront payment V(0) can of course be used to *actually* fund the deposit, but the applicable interest rate is still $r_b(t)$ as it could otherwise be used to pay down B's debt.

Because of the deposit from the liability side, the only account potentially subject to default settlement is the borrowing account $N_t$, which has balance zero under perfect replication. Suppose that the unit notional swap has a cumulative dividend process $D(t)$, the pre-default financing equation is written as follows

$$d\Delta_t(V^* + dV^*) - \Delta_t dD + dD + dL - r_c L^+ dt + r_b L^- dt - dL^s + r_L L^s dt = 0$$

Since swap #1 is a single period swap, $dD=0$ for $t<T$. If the interest rate $\rho$ governing the swap payoff is modeled after a diffusion process under an equivalent martingale measure Q in a properly defined probability space $\{\Omega, \mathcal{G}, P\}$, $d\rho = adt + bdW$, $V^*$ solves the following partial differential equation,

$$\frac{\partial V^*}{\partial t} + a\frac{\partial V^*}{\partial \rho} + \tfrac{1}{2}b^2 \frac{\partial^2 V^*}{\partial \rho^2} - r_L V^* = 0$$

For simplicity, we have assumed all short rates are a function of $\rho$ (for example, deterministic spread to $\rho$). Noting the above PDE and applying Ito's lemma to $V_t$ and $V^*$ lead to

$$(\frac{\partial V}{\partial t} + a\frac{\partial V}{\partial \rho} + \tfrac{1}{2}b^2 \frac{\partial^2 V}{\partial \rho^2} + r_b V^- - r_c V^+)dt + b\frac{\partial V}{\partial \rho}dW - b\Delta\frac{\partial V^*}{\partial \rho}dW = 0$$

Assume swap delta hedge $\Delta = \frac{\partial V}{\partial \rho} / \frac{\partial V^*}{\partial \rho}$, and set $dt$ term to zero, we arrive at,

$$\frac{\partial V}{\partial t} + a\frac{\partial V}{\partial \rho} + \tfrac{1}{2}b^2 \frac{\partial^2 V}{\partial \rho^2} + r_b V^- - r_c V^+ = 0$$

This is the same PDE for uncollateralized options when drift and diffusion coefficients $a$ and $b$ are replaced with those of a stock's lognormal price process in the risk neutral world. In fact, the extended option PDE can be derived alternatively by hedging an uncollateralized option with a CCP or exchange traded option and simply changing the dynamics of the short rate to that of the stock price.

The terminal boundary condition is $V_T = V^*_T = D_T$, where $D_T$ is payoff of the swaplet at maturity $T$. As $V^*_T$ is simply the terminal swap payment, enforcing the terminal condition guarantees that the hedge scheme produces the same swap cashflow as t approaches T. In fact, from $V_T = V^*_T$, differentiate with respect to $\rho$, we have $\frac{\partial V_T}{\partial \rho} = \frac{\partial V^*_T}{\partial \rho}$, i.e. $\Delta_T = 1$, the swap hedge ratio is one on swap payment date.

The one period uncollateralized swap is thus fully replicated, in both interest rate risk and cash flow, by dynamically trading a same term, same fixed rate swap with a CCP member firm. In other words, the uncollateralized swap (swap #1) is attained from a CCP swap delta hedge and a deposit under the liability-side pricing principle.

Finally by means of Feynman-Kac theorem, we have the familiar expected discounted payoff formula,

$$V(t) = E_t[e^{-\int_t^T r_e du} D_T],$$

Zero coupon bond pricing can be treated as a special case by setting $D_T$ to 1. If the bond is issued by party C, then $r_e = r_c$, or $r_e = r_b$ if by party B. The zero coupon bond is always priced at its issuer's cash bond yield as the case should be.

The framework outlined above works for caps and floors, and can be easily extended to price foreign exchange rate (fx) swaps and other rates and fx derivatives.

## 3. Swap Valuation with Coherent CVA and FVA

Heuristically, a swap of regular payment dates can be decomposed into a series or a portfolio of swaplets, each being replicated, in both risk and cash flow, by an identical CCP swaplet. At any given time t, each swaplet has its own hedge ratio, so the resultant portfolio of CCP swaplets is of non-uniform notionals on each payment date, therefore not a typical CCP swap. But this can be easily mapped into a portfolio of CCP swaps having stacked maturities and different notionals determined via a boot-stripping procedure. In the end, the uncollateralized swap is attained by a portfolio of CCP swaps having different tenors and different notionals. The liability-side market funding, however, is provided at the portfolio level, on a netted basis.

Let $T_i$ denote the $i$-th payment time of the swap, $\delta_i$ swap payment at $T_i$, $V_i$ the fair value of the $i$-th swaplet facing party C, we can write,

$$V(t) = \sum_{t < T_i} V_i(t),$$
$$V_i(t) = E_t^Q[e(-\int_t^{T_i} r_e du)\delta_i],$$

As $t$ approaches $T_i$, $V_i(t)$ approaches the cash payment itself with probability 1, if all short rates involved are continuous. In practice this is guaranteed as swaps are almost exclusively in-arrear. If we denote $V^*_i(t)$ as the OIS discounted CCP swaplet, then we always have $V_i(T_i) = V^*_i(T_i)$.

From the PDE perspective, between payment dates, the same PDE would apply to each swaplet and thus to the swap as a portfolio of swaplets. On a payment date, one swaplet will drop out, creating a discontinuity for the swap fair value. Assuming a properly regulated swap payment or dividend process, the solution to the PDE can be summarized in the following proposition.

**Proposition 1 (Liability-Side Pricing of Uncollateralized Swaps)**: A bilateral cleared, uncollateralized swap's no-arbitrage fair value is the expected risky discount of the swap's net cashflow under the risk neutral measure Q, $V(t) = E_t^Q[\int_t^T e^{-\int_t^s r_e du} dD_s]$, where $D_t$ is the dividend process and $r_e$ is the discount rate switch, $r_e(t) = r_b(t)I(V(t) < 0) + r_c(t)I(V(t) \geq 0)$.

The expectation taken under the risk neutral measure Q shows swap valuation by means of risky discounting of the cash flow. The discount rate, however, is switching between the two parties based on the swap being a receivable or a payable locally in time. Duffie and Huang (1996) encounter a similar curve switching where the discount rate is $r + s_A 1(V<0) + s_B 1(V>=0)$, $s_A$ and $s_B$ are counterparty A's and B's CDS rates respectively, $s_A = (1-R_A)*h_A$, $s_B = (1-R_B)*h_B$, $R_A$ and $R_B$ A and B's recovery rates on pre-default market values with $h_A$ and $h_B$ as hazard rates in the risk-neutral world. The difference is that their switching is on the CDS rates while here it is on the cash rate, or full yield spreads when the risk free rate r is taken out, i.e., with the addition of the funding basis on top of the CDS rates.

We could in fact incorporate a carrying cost (or benefit if negative npv) as a fraction of swap npv in Duffie and Huang's model, with the same technical rigor, to arrive at the same swap valuation formulae (see Appendix for a proof of Proposition 1). The approach taken there is the classic reduced form, risk-neutral pricing approach for credit derivatives. In the above derivation, the risk neutral measure is only used to establish the pricing of fully collateralized or CCP swaps which the industry has consensus of applying the OIS curve as the risk free rate. The uncollateralized swap then is fully replicated by continuous trading in the CCP swap, without reference to the risk neutral default intensities and recovery rates. In typical credit derivatives pricing, the intensity or hazard rate is calibrated to CDS market, while the recovery rate is not market observable and set at some historical mean or by convention, for instance, 40% for investment grade credits. Note that a bond yield curve has no specification and no need of a recovery rate[5]. The liability-side pricing of uncollateralized swaps thus can be seen as an application or extension of the risk neutral pricing theory.

In practice the risk neutral measure is calibrated to a series of standard market swaps, so the pricing of the uncollateralized swap connects directly to these market swaps. A popular interest rate risk hedge scheme is to use a product's sensitivities to swap rates that constitutes the discount curve. The same scheme applies here. Specifically we could seek sensitivities of $V(t)$ with respect to the OIS discounted swap curve and determine hedge ratios in each tenor. The advantage of such a scheme is that it does not depend on the factors that drive the curve. A bilateral uncollateralized swap is fully hedged by trading in CCP swaplets.

$$V(t) = \sum_i V_i = \sum_i \Delta_i V_i^*,$$

$$\Delta_i = \frac{\partial V_i}{\partial \rho} / \frac{\partial V_i^*}{\partial \rho},$$

$$V_i^* = E_t^Q [e(-\int_t^{Ti} r du) \delta_i],$$

where $V_i$ are the risky swaplet fair price. These swaplet hedge ratios can then be reverse boot-strapped into regular CCP swaps with staggered tenors.

The total counterparty risk adjustment (CRA) to the counterparty default free fair value $V^*$ can be formulated exactly,

---

[5] Reduced form models are distinguished in their specification by the use of an exogenously given recovery rate, commonly as a fraction of the face value or pre-default market value. The credit derivative markets responded with some not too successful initial efforts of developing a recovery swap market which basically disappeared after the financial crisis. Time to return to a bond like, recovery-rate-less model?

$$V^* - V = E_t^Q[\int_t^T (r_e - r)V^*(s)\exp(-\int_t^s r_e du)ds]$$

CRA is decomposed into bilateral CVA and FVA, in accordance to the decomposition of the credit spread into a default risk (CDS) component and a liquidity or funding basis.

**4. A Least-Square Regression/Simulation Liability-side Pricing Model**

The liability-side pricing principle nicely integrates derivatives' market risk and counterparty credit risk in one formula, or one model. Except for pure asset or liability products such as a cap or floor which allow natural decoupling of the effective discount rate with the fair value, the expectation formula is of limited practical use as the effective rate depends on local fair value of the swap and numerical solutions have to be sought.

On a trade level or option trading strategy that involves longs and shorts of options on the same underlying, a finite difference (FD) method is developed to solve the PDE for the fair value of the trade (Lou 2015b). As will be shown in this section, similar FD method can be developed for swaps or portfolio of swaps and rate derivatives when the underlying factors are very few. A derivatives netting set is an intrinsically high-dimension portfolio, so pricing, CRA, CVA and FVA can only be done by Monte Carlo (MC) simulation. A particular challenge has been the often recursive nature of valuation adjustments. In the liability-side pricing model, the recursion happens only through the applicable discount factor and is different from other models, for example Burgard and Kjaer 2011, where the coupling binds the unknown fair value and its adjustment through some exposure amounts to be risk-free discounted.

The aim of this section is to provide techniques to decouple the recursion and a starter implementation in both MC and FD where FD could be used to shed lights on funding cost's impact on swaps pricing and valuation and to verify MC implementation. For illustrative purposes, the FD and MC implementations and computational results are based on a one-factor short rate model, either mixed-normal-lognormal model or the classic Black-Karasinski model.

**4.1 Short rate models**

CCP swaps are priced in dual curve settings where both the OIS curve and the LIBOR curve (or the LIBOR/OIS spread curve) are modeled. For illustrative purposes, we choose a one-factor LIBOR short rate model and leaves the risk free rate (OIS) $r$ at a deterministic spread off the LIBOR short rate, $\rho - r >= 0$, where $\rho$ is the LIBOR short rate. As our focus is counterparty credit and funding risk, such a simple model could be adequate. Multi-factor short rate, HJM, and market models with joint modeling of the OIS rate and the LIBOR-OIS spread can be easily developed within the same simulation framework of the liability-side pricing model.

In the current low rate environment, a normal short rate model such as the Hull-White model has a significant part of short rate distribution in the negative territory and CIR model has difficulty calibrating to the markets. A recent improvement is to fit the volatility of the short rate as a function of the rate itself (DeGuillaume et al 2013). It is found that the short rate could be cut into three regions from zero rate to certain maximum rate. In the beginning and end regions, the rate's volatility is approximately linear to rate, while in the middle region, the volatility is about constant. The short rate is therefore locally lognormal except in the middle where it is normal (mixed-normal-lognormal model or mixed model in short). Specifically write $dr_t = a(\theta - r)dt + \sigma(r)dW$, where the volatility function is defined as $\sigma(r) = \frac{r}{0.015}\sigma_2$, for r<0.015, $\sigma(r) = \sigma_2$ for $0.015 \leq r < 0.06$, and $\sigma(r) = \frac{r}{0.06}\sigma_2$ for $r \geq 0.06$, with parameters estimation on historical data *σ2=1.05%, a=0.05, θ=0.044* (Hull et al 2014). As a comparison, a mean-reverting, lognormal model or Black-Karasinski (B-K) model is also implemented, $dx_t = \kappa(\mu - x)dt + \sigma dW$, $r_t = \exp(x_t)$, where *κ, µ, σ* are positive constants.

**4.2. Decoupling techniques**

In the industry, CVA is computed via Monte Carlo simulations. The main obstacle of computing the LSP fair value is the discount rate switch that depends on the local fair value, i.e., the fair value equation is recursive. As the recursion only happens on the discounter factor, the coupling is a weak one, allowing approximations to take place.

Consider a small time step *[t, s), s=t+dt*. *V(t)* is right-continuous-left-limit (RCLL), with its discontinuity reflecting discrete swap payment at time *t*. If *dDt* is the payment paid at time *s*, *V(s_)* is the left limit, then *V(s_)=V(s)+dDt*. Applying conditional expectation at time *t* and *s* iteratively, we have the following one step valuation equation,

$$V(t) = E_t^Q[e^{-\int_t^s r_e du} V(s\_)],$$
$$r_e = r_c + (r_b - r_c)I(V(u) < 0)$$

As *dt* is small, and *V(t)* is continuous in *[t, s)*, our first approximation is to apply the indictor function and the rates at the beginning of the period, i.e., time *t*. The discount factor can then be taken out of the expectation to yield,

$$V(t) = e^{-dt(r_c(t) + (r_b(t) - r_c(t))I(V(t)<0))} E_t^Q[V(s\_)],$$

Notice that the discount factor is positive so *V(t)* has the same sign as $E_t^Q[V(s\_)]$, i.e., $I(V(t) < 0) = I(E_t^Q[V(s\_)] < 0)$. Now *V(t)* is decoupled and can be computed directly from $E_t^Q[V(s\_)]$.

If we fix the effective rate at the period end instead, the expectation can be evaluated directly, knowing the distribution of rates at s and *V(s_)*. Since rates are only mildly diffusive, a further approximation is to use the rates at *t* but leave the switch at *s*. The following formula then results,

$$V(t) = (1 - r_c(t)dt)E_t^Q[V(s\_)] + dt(r_b(t) - r_c(t))E_t^Q[V^-(s\_)],$$

where $V^-$ is the negative part of *V*. A more generic approximation involving a weighted average of the effective rates at the beginning and end of the period can be developed similarly, if necessary.

These one-step decoupling techniques can be used when local simulation is conducted or on a lattice or tree.

**4.3. Finite Difference Solver**

A Crank-Nicholson FD scheme (Lou 2015b) is developed to cope with the free asset-liability boundary problem of an option trading strategy, where a projected successive over relaxation algorithm (PSOR) is used to handle the rate switch.

For swaps, the discrete cash flows need special care. In between payment dates, the PDE applies while on payment dates, swap npv will jump in the amount of the payment. Duffie and Huang (1996) proposes a finite difference scheme to capture a swap's reset and payment-in-arrears feature. The short rate on reset dates is treated as an auxiliary state variable which is fixed until the next reset. This enables the finite difference solver to run largely as a one-dimensional scheme, although limited in that it can only handle non-overlapping reset periods. It is not clear however how the rate switch is treated. Our FD solver incorporates discrete swap payments and resorts to simple iteration to make sure that the switch is compatible with the fair value.

In order to capture the periodic floating rate payments, LIBOR rates have to be calculated. Unlike a CIR model where analytic formula exists for the price of zero coupon bonds, which can be used to convert the LIBOR short rate to 3 month LIBOR rate, the B-K model does not have an analytically tractable solution. An approximation technique developed by Xu (2014) is utilized to calculate the 3 month LIBOR rate given a grid of the LIBOR short rates. For the mixed-normal-lognormal model, we use the same FD solver to calculate 3 month zero coupon bond's price and (pre)compute LIBOR rates for the short rate grid.

**4.4. LSP Simulation Model**

The discount rate switch is a binary category same as the early exercise condition of American option is: the latter compares a continuation value with a payoff value to determine whether the option is exercised or not, while the former compares the continuation value with value zero to determine which discount rate shall be used. Motivated by Longstaff and Schwartz (2001), we adopt a similar regression scheme to resolve the switch along with swap valuation.

We set up a discrete time epoch $t_0 < t_1 < t_2 < ... < t_K = T$. Introduce a switching matrix on all paths and time steps *B(j,k)*, with value 0 indicating using $r_b$ and *1* $r_c$, where j is the path index, and *k =0, 1, 2, …, K* is the time step index. Starting from maturity of the swap *T*, we know the pathwise net swap payments. Roll backward on the path from $t_K$ to $t_{K-1}$ by discounting those positive with $r_c$, or those negative with $r_b$, to arrive at path value *V(j,K-1)*. Next, apply least-square regression to *V(ω,K-1)* on given set of basis functions of short rates, *ω* representing all paths. The fitted value of the regression *U(ω,K-1)* is then used to set the switching matrix at $t_{K-1}$ *B(j,K-1)*. Now we add swap payment at $t_{K-1}$ to *V(ω,K-1)* to arrive at *V(j, $t_{K-1}$-)*, then proceed with discounting *V(j, $t_{K-1}$-)* with the discount rate decided based on the switching matrix (rather than the sign of *V(ω,K-1)*).

Specifically, the following procedure is developed as a starter implementation of coherent CVA and FVA computation,
1.) Simulate the LIBOR short rates on a path,
2.) Compute all relevant rates (synthetic rates and bond rates) as deterministic spreads to the short rate,
3.) Starting from the longest maturity, determine swap payments as terminal fair value V(T), and U(T),
4.) Initialize switching matrix B(T) to zero, and set to 1 where U(T) are positive,
5.) At time step t+dt, Choose discount rates $r_c$ if U is switched on (with value 1), otherwise $r_b$,
6.) Discount V(t+dt) one time step to get V(t),
7.) Perform least-square regression of V(t) with chosen basis functions,
8.) Reval local fair value from the regression function to get U(t),
9.) Update switching matrix B(t) based on U(t),
10.) Add swap payments on t to V(t) and repeat.
11.) Take the mean of V(0) and compute standard error.

With regards to the basis functions, there are a good number of choices. For the results presented herein, a second order Laguerre polynomial is used. Higher orders and use of simple polynomial of order 2 or higher all produce basically the same results, although implementations shall experiment with their specific netting sets. Taking out the regression, the procedure reduces to a brute force MC simulation, which will be shown to be inaccurate.

The incremental decomposition of the total counterparty risk adjustment into CVA/DVA, FCA/FDA necessitates revaluation of swaps and derivatives under a different set of discounting curves. From Lou (2015a), if we write the portfolio pricer as $V(f_b, f_c)$ where $f_b$, $f_c$ are placeholders for the discount curves of party B and C respectively, then we have the following quantities linked to the pricer under different sets of curves,

$$U = V(r, r) - V(r_b, r_c),$$
$$CVA = V(r, r) - V(r, \tilde{r}_c),$$
$$DVA = -V(r, \tilde{r}_c) + V(\tilde{r}_b, \tilde{r}_c),$$
$$FCA = V(\tilde{r}_b, \tilde{r}_c) - V(\tilde{r}_b, r_c),$$
$$FDA = -V(\tilde{r}_b, r_c) + V(r_b, r_c)$$

The effective rate switch is still on the fair value to simplify the computation. This enables the simulation engine to roll back on time step simultaneously on these pairs of discount rates and compute all shifted prices at the same time.

Multi-factor short rate, HJM, and market models with joint modeling of the OIS rate and the LIBOR-OIS spread can be easily accommodated within the simulation procedure. A FD solver incorporating discrete swap payments and iterative procedure to handle the rate switch is developed to verify the LSP simulation model.

## 5. Computational results

For results presented below, the mixed-normal-lognormal model has the following parameters, a=0.21, $\sigma_2$=2.52%, $\rho_0$=0.18%, to calibrate to 3 month LIBOR of 0.2887%, 5 year swap rate at 172.666 bp and 5 year ATM strike cap at yield value[6] of 73 bp. B-K model parameters are calibrated to the same LIBOR rate, swap rate and cap prices with μ=0.044, κ=0.2809, σ=0.8273, $\rho_0$=0.25%. The LIBOR short rate and the risk-free rate spread is assumed at recent 3 month Libor-OIS average spread of 13 bp. The full cash funding curves are assumed to be a deterministic spread above the LIBOR short rate or the risk-free short rate. For 10 year results, both models are calibrated to 10 y swap rate of 235.87 bp and 10 y cap with strike at ATM swap rate is 86.83 bp.

### 5.1. Regression/Simulation vs FD

To test the LSP simulation model, Table 1 lists a 10y ATM swap npv (in yield value) under the B-K model and the mixed model. Party B is set at LIBOR flat while C has a spread to B at 250 bp, 500 bp, and 1000 bp. The number of paths is 100,000, resulting standard error of swap npv is 0.1 bp. The time steps for both MC and FD are 0.0125 years. As we can see, the difference (last row) between the FD and MC (3$^{rd}$ row, labeled LS-MC) is about few hundredths of a basis point. A brute force simulation without least-square regression (4$^{th}$ row), however, has significant errors. With the mixed-normal-lognormal model, when C's spread to B is 250 bp, the error is 3.4 bp, which grows to 9.6 bp as C's spread widens to 1000 bp. This shows that handling the rate switch with regression is essential for an accurate simulation of LSP pricing and CVA and FVA computation.

Table 1. LSP simulation model verification with FD model for a 10y ATM swap under both the B-K model and the mixed model.

|        | B-K Model |         |         | Mixed Model |          |          |
|--------|-----------|---------|---------|-------------|----------|----------|
| C-sprd | 1000      | 500     | 250     | 1000        | 500      | 250      |
| FD     | -29.3137  | -17.2979| -9.5196 | -23.4533    | -13.4788 | -7.2889  |
| LS-MC  | -29.3227  | -17.311 | -9.5297 | -23.4389    | -13.4566 | -7.2587  |
| MC     | -37.3454  | -22.397 | -12.3747| -33.0324    | -19.5451 | -10.6807 |
| Diff   | -0.009    | -0.0131 | -0.0101 | 0.0144      | 0.0222   | 0.0302   |

### 5.2. Switch on net cashflow or net pv?

Hubner (2001) popularizes a method of cash flow discounting first by segregating net cash inflow and net cash outflow and applies counterparty's credit curve to net inflow and own credit curve to outflow respectively to get present values. The inflow pv and outflow pv are netted to arrive at the swap's npv (net present value). This method is intuitive at the time and is similar to the liability-side pricing in that the discount curve switches between the counterparties' credit curves. The difference is obvious as it switches on the sign of the net swap cash flow while LSP is on the sign of the swap npv.

---

[6] A yield value is npv normalized by notional and annuity of the same term swap, in basis points.

This type of aggregation and discounting scheme is obviously flawed in theory. Consider, for example, a hypothetic swap of only one exchange of a payment at *T* and a receipt of the same amount at *T-dT*. If *dT* is zero, these two payments net out to zero so that the swap has zero npv. Now with a very small but non-zero *dT*, the receipt will be discounted at the counterparty's rate $r_c$ while the payment will be at $r_b$. Obviously when *dT* approaches zero, npv of the swap is not continuous and the gap[7] is at the mercy of the relative credit spread $r_c-r_b$.

As the method is quite popular with practitioners, it is interesting to see how much approximation error it could induce. A payer (paying fixed, receiving float) swap decomposes into a long cap and a short floor. Long cap is the net inflow while the floor is the net outflow, so Hubner's approach is equivalent to pricing the swap as a synthetic swap – a portfolio of long cap and short floor discounted at $r_c$ and $r_b$ respectively. Its price deviation from the swap npv reflects the portfolio netting effect. Table 2 shows the ATM payer swap, long cap, and short floor's npv in yield value (bp) under the mixed normal and lognormal model. As C's spread widens while keeping B's at the risk free rate, the swap npv turns increasingly negative, reflecting the increased counterparty default and funding risk. At 250 bp, for instance, yield value of the swap is -3.3 bp. The floor yield remains the same (as is party B's liability) but the cap yield drops to 67.4 bps so that cap minus floor is -6.2 bps, 2.9 bp difference from the swap npv solved directly from the FD solver, as shown in the last column. The netting effect is very significant, in the same magnitude as the swap npv.

Table 2. ATM 5y swap (at 172.67 bps) priced under mixed-normal-lognormal model as compared to cash flow discounting of cap and floor.

| C-sprd | swap-yld | cap – floor | cap yld | floor yld | Diff |
|---|---|---|---|---|---|
| 0 | 0.1 | 0.1 | 73.7 | -73.6 | 0 |
| 125 | -1.6 | -3.1 | 70.5 | -73.6 | 1.5 |
| 250 | -3.3 | -6.2 | 67.4 | -73.6 | 2.9 |
| 500 | -6.5 | -11.9 | 61.7 | -73.6 | 5.4 |
| 1000 | -12.2 | -21.8 | 51.8 | -73.6 | 9.6 |
| 2000 | -21.3 | -36.8 | 36.8 | -73.6 | 15.5 |

Deep in-the-money (ITM) swap is interesting in that it involves upfront funding. Table 3 shows an ITM receiver swap with strike at 500 bps (ATM swap rate at 172.67 bp), under the B-K model. Swap npv is more sensitive to C's spread widening. Swap npv yield, for example, drops 17.8 bp from 329.5 to 311.7 when C's spread at 250 bps, compared to 3.3 bp drop for the ATM swap as shown in Table 2. The netting effect, while in the same magnitude as the ATM swap netting effect, is less pronounced comparing to the deep ITM npv.

---

[7] As a numerical example under the BK FD model, where $r_b$ is at LIBOR while $r_c$ is 125 bps wider, applying Hubner's discounting of cash flow at T=5 years would result in a npv of -5.725% of notional, while solving the LSP PDE would only give -0.223%. At 250 bps wider, Hubner's results goes out further to -10.947%, while FD solution is little changed at -0.281%.

Table 3. Deep ITM 5y swap (at 500 bps) priced under the Black-Karasinski model as compared to cash flow discounting of cap and floor.

| C-sprd | Swap npv | flr-cap | cap yld | floor yld | BK-Diff |
|---|---|---|---|---|---|
| 0 | 329.5 | 329.5 | -27.3 | 356.8 | 0 |
| 125 | 320.4 | 319.5 | -27.3 | 346.8 | 0.9 |
| 250 | 311.7 | 309.8 | -27.3 | 337.1 | 1.9 |
| 500 | 295.2 | 291.6 | -27.3 | 318.9 | 3.6 |
| 1000 | 265.8 | 259.1 | -27.3 | 286.4 | 6.7 |
| 2000 | 218.5 | 207 | -27.3 | 234.3 | 11.5 |

Although not accurate, switching on net cashflow allows a bank's existing CVA infrastructure to migrate easily. For CVA and DVA based on expected positive and negative future exposure (EPE, ENE) simulation, the following is an approximation to coherent CVA and FVA,

$$CVA = E_t^Q[\int_t^T (r_c - r - \mu_c) EPE(s) \exp(-\int_t^s r_c du) ds],$$

$$CFA = E_t^Q[\int_t^T \mu_c EPE(s) \exp(-\int_t^s r_c du) ds],$$

$$DVA = E_t^Q[\int_t^T (r_b - r - \mu_b) ENE(s) \exp(-\int_t^s r_b du) ds],$$

$$DFA = E_t^Q[\int_t^T \mu_b ENE(s) \exp(-\int_t^s r_b du) ds]$$

where $\mu_b$ and $\mu_c$ are the funding basis of B and C respectively.

**5.3. Swap rate and CCP swap hedge ratios**

Swap pricing aims to determine a swap rate such that the swap npv is zero. Figure 1 shows that the swap rate gradually declines from the risk free ATM rate of 172.7 bps to 154.8 bps when party C's credit spread widens to 10%. The ATM swap rate initially declines at a rate of 1.8 basis point per 100 bps of C's credit spread widening, reflecting elevated sensitivity in current interest rate environment which is very different from 1990's when both spot and historical interest rates are relatively high and Duffie and Huang (1996) report about 1 basis point increase in swap rate, given a typical CIR model with historical mean interest rate at 10%.

Swap delta or sensitivity to CCP swaps follows a similar pattern to the swap rate bid/ask spread. Figure 2 shows a decline from 1 to 0.85 and 0.81 for 5y ATM swap and ITM 5% receiver swap respectively. Initial rate of change is 2% drop per 100 bp widening under the mixed-normal-lognormal model. The hedge ratio of an OTM 5% payer swap is steady, only declining to 0.98, as there is little counterparty adjustment for party B is at LIBOR flat.

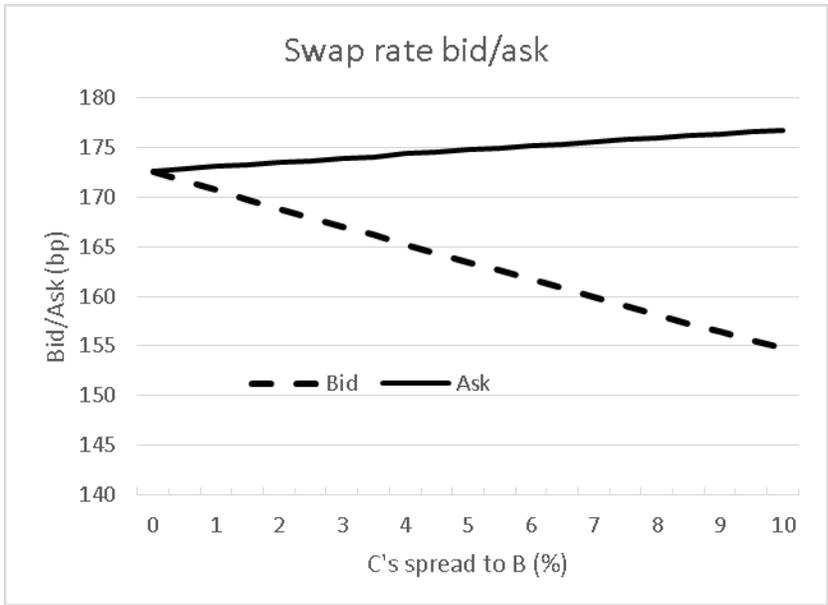

Figure 1. ATM swap rate bid and ask as counterparty's credit spread widens.

The switching effective rate naturally induces swap bid/ask spread. When the cash curves $r_b$ and $r_c$ are not identical, the PDE is position asymmetric, meaning that party B will see a receiver swap's npv different from a payer's npv. Consequently pricing the swap on the payer side will lead to a rate different from the rate on the receiver side, thus creating a bid/ask spread, in the same spirit as an option market maker's funding cost creating an option bid and ask spread (Lou 2015b).

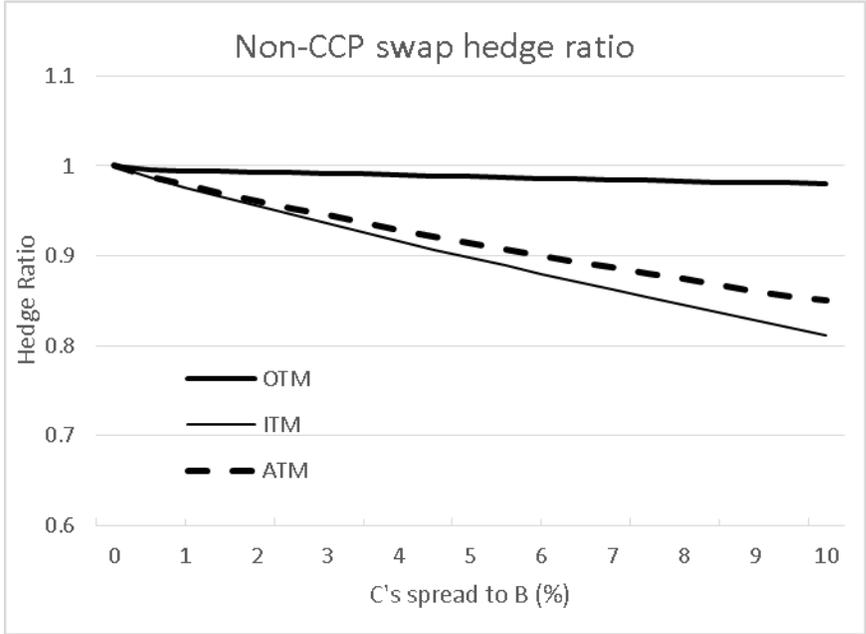

Figure 2. Uncollateralized or Non-CCP swap hedge ratio decreases as party C's spread widens to LIBOR where B is at.

Fixing B's credit spread to OIS at 125 bps and C's relative spread to B at 125 bps leads to a bid of 167.27 bps and ask of 170.22 bps, a spread of 3 bps. The bid and ask further diverge as C's relative spread increases, resulting in wider bid/ask spread. The bid/ask spread is 22.44 bps, for example, when C has a spread of 1000 bps,

**5.4. Swap CRA and decomposition results**

To demonstrate, we set Party B's CDS (zero recovery rate) short rate to 75 bps and its funding basis 50 bps on top of LIBOR, approximated single 'A' rated. C's spread to LIBOR varies from 37.5 to 1000 bp, roughly reflecting credit rating range of "AAA/AA+" to "B". C's funding basis are 15, 30, 50 bp for the first three 'rating's and 80 bp for the rest. All values in Table 4 and 5 are yield values in bp computed by the LSP simulation model.

For the ATM receiver swap, DVA and DFA are larger, intuitively due to greater negative exposure.

Table 4. 10y ATM payer swap valuation adjustments under the mixed model with a single 'A' rated dealer facing a counterparty of hypothetic 'AAA/AA+' to 'B' rating.

| C-Libor | "Rating" | NPV | CRA | CVA | DVA | CFA | DFA |
|---|---|---|---|---|---|---|---|
| 37.5 | AAA/AA+ | -0.45 | 0.84 | 1.13 | 0.49 | 0.47 | 0.27 |
| 75 | AA/AA- | -1.58 | 1.97 | 1.83 | 0.5 | 0.91 | 0.27 |
| 125 | A | -3.05 | 3.44 | 2.75 | 0.5 | 1.47 | 0.28 |
| 250 | BBB | -6.5 | 6.87 | 5.52 | 0.53 | 2.18 | 0.3 |
| 500 | BB | -12.6 | 12.98 | 12.05 | 0.59 | 1.86 | 0.34 |
| 1000 | B | -22.37 | 22.76 | 22.52 | 0.74 | 1.4 | 0.42 |

Table 5. 10y ITM 5% receiver swap valuation adjustments under the mixed model with a single 'A' rated dealer facing a counterparty of hypothetic 'AAA/AA+' to 'B' rating.

| C-Libor | "Rating" | NPV | CRA | CVA | DVA | CFA | DFA |
|---|---|---|---|---|---|---|---|
| 37.5 | AAA/AA+ | 260.52 | 4.9 | 3.79 | 0.3 | 1.58 | 0.17 |
| 75 | AA/AA- | 256.64 | 8.79 | 6.15 | 0.3 | 3.1 | 0.16 |
| 125 | A | 251.59 | 13.84 | 9.25 | 0.3 | 5.05 | 0.16 |
| 250 | BBB | 239.61 | 25.82 | 18.7 | 0.29 | 7.56 | 0.15 |
| 500 | BB | 218.12 | 47.31 | 41.17 | 0.26 | 6.54 | 0.14 |
| 1000 | B | 183.29 | 82.14 | 77.53 | 0.22 | 4.95 | 0.12 |

The CRA of the ITM swap when C's spread to LIBOR is 47.3141 solved from the Crank-Nicolson FD solver, compared to 47.3001 bp from simulation. This again shows the satisfactory accuracy of the LSP simulation model.

While the results are given on a single trade or a simple portfolio, LSP pricing applies at the counterparty or firm level to allow netting of funding in a whole derivatives portfolio.

## 6. Conclusions

By constructing a dynamic CCP swap hedge, an uncollateralized swap is shown to be fully replicated, allowing no-arbitrage pricing of swaps subject to counterparty default and funding risk. The open IR01 of a back-to-back, static hedge is plugged in and there is no need of market risk capital add-on. Swap valuation follows the same liability-side pricing principle established for stock options, and the total counterparty risk adjustment is the cost of financing the CCP swap mark-to-market from the liability-side counterparty's perspective, leading to coherent definition of bilateral CVA and FVA.

Computation of fair value and CVA and FVA involves solving a recursive expectation due to LSP model's distinct discount rate switch. A least-square regression/simulation scheme is devised to compute CVA, DVA, CFA (FCA), and DFA (FBA) simultaneously, and is found indispensable as brute force MC simulation produces large errors. Preliminary results from calibrated mixed-normal-lognormal short rate and Black-Karasinski models show that for moderate credit spread, swap rate can change by about 2 bp in current rate environment and that the swap hedge ratio deviates away from 1-to-1 by about 2% with each 100 bp difference in counterparties' credit spreads. LSP naturally applies to counterparty netting or firm funding set level, so long as an implementation could accommodate other rate models and products, a topic to be explored in the future for an industry strength implementation of coherent CVA and FVA.

# Appendix: Proof of Proposition 1

Following Duffie and Huang (1996)'s notations, the cumulative dividend process of the swap for counterparty A under the liability side market funding principle (Lou 2015a) is defined by,

$$X(t) = \int_0^t (1-H_u)(dD_u - S_u(1(S_u \geq 0)(r_b - r) + 1(S_u < 0)(r_c - r))du) + \int_0^t S_{u-}dH_u,$$

where $H_t$ is same as our $\Gamma$, $S_t$ is the market value of the swap, $D_t$ is the natural swap cumulative payment process without consideration of counterparty default. Note here we are only paying the spread above the risk free rate out of the swap cash flow as the deposit would earn the risk free rate return.

Under the equivalent martingale pricing measure, the swap market value is

$$S_t = E_Q[\int_t^T \exp(-\int_t^s r du) dX_s \mid F_t],$$

Substituting $X_t$ into the above, we obtain

$$S_t \exp(-\int_0^t r du) = E_Q[\int_t^T \exp(-\int_0^s r du)[1\{s < \tau\}(dD_s - S_s \mu ds) + S_{s-}dH_s] \mid F_t],$$

where $\mu_t = 1(S_t \geq 0)(r_B - r) + 1(S_t < 0)(r_A - r)$ is the effective short rate spread.

Noting that $dH_t = 1\{t < \tau\}(h_t^A + h_t^B)dt + dM_t$, where h are the hazard rates of party A and B respectively, M is a Q-martingale per Doob-Meyer decomposition. Differentiate the above to arrive at,

$$dS_t = [R_t - h_t^A - h_t^B]S_t dt - 1\{s < \tau\}dD_t + d\hat{m}_t,$$

where $R_t = r_t + \mu_t$. The rest follows Appendix A (Duffie and Huang 1996). This proof shows that the same pricing measure for the CCP swaps is unaltered in order to price the uncollateralized swaps, contrary to Kenyon and Green's (2014) argument.